\documentclass[preprint,12pt,authoryear]{elsarticle}

\usepackage{amsmath,amssymb,amsfonts}

\usepackage[utf8]{inputenc}
\usepackage[T1]{fontenc}
\usepackage[english]{babel}
\usepackage[hidelinks, unicode]{hyperref}
\usepackage[table,xcdraw]{xcolor}
\usepackage{url}
\usepackage{csquotes}
\makeatletter
\let\c@author\relax
\makeatother
\usepackage[style=ieee, backend=biber, url=false, doi=true, eprint=false, isbn=false, date=year, 
dashed=false
]{biblatex}

\usepackage{xfrac}
\usepackage{nidanfloat}
\usepackage{orcidlink}
\usepackage{multirow}

\AtEveryBibitem{%
  \clearfield{note}%
}

\journal{Biomedical Signal Processing and Control}

\begin{document}

\begin{frontmatter}



\title{Deep comparisons of Neural Networks from the EEGNet family}


\author[inst1,inst2,inst3]{Csaba M\'arton K\"oll\H{o}d \orcidlink{0000-0003-3817-6709}}

\affiliation[inst1]{organization={Roska Tamás Doctoral School of Sciences and Technology},
            city={Budapest},
            country={Hungary}}

\author[inst1,inst2,inst3]{Andr\'as Adolf \orcidlink{0000-0001-7855-1492}}
\author[inst2,inst3]{Gergely M\'arton \orcidlink{0000-0003-4359-0111}}
\author[inst2,inst3]{Istv\'an Ulbert \orcidlink{0000-0001-9941-9159}}

\affiliation[inst2]{organization={Faculty of Information Technology and Bionics, P\'azm\'any P\'eter Catholic University},
            city={Budapest},
            country={Hungary}}

\affiliation[inst3]{organization={Cognitive Neuroscience and Psychology, Research Centre for Natural Sciences},
            city={Budapest},
            country={Hungary}}


\begin{abstract}
Most of the Brain-Computer Interface (BCI) publications, which propose artificial neural networks for Motor Imagery (MI) Electroencephalography (EEG) signal classification, are presented using one of the BCI Competition datasets. However, these databases contain MI EEG data from less than or equal to 10 subjects  . In addition, these algorithms usually include only bandpass filtering to reduce noise and increase signal quality. In this article, we compared 5 well-known neural networks (Shallow ConvNet, Deep ConvNet, EEGNet, EEGNet Fusion, MI-EEGNet) using open-access databases with many subjects next to the BCI Competition 4 2a dataset to acquire statistically significant results. We removed artifacts from the EEG using the FASTER algorithm as a signal processing step. Moreover, we investigated whether transfer learning can further improve the classification results on artifact filtered data. We aimed to rank the neural networks; therefore, next to the classification accuracy, we introduced two additional metrics: the accuracy improvement from chance level and the effect of transfer learning. The former can be used with different class-numbered databases, while the latter can highlight neural networks with sufficient generalization abilities. Our metrics showed that the researchers should not avoid Shallow ConvNet and Deep ConvNet because they can perform better than the later published ones from the EEGNet family.
\end{abstract}



\begin{keyword}
BCI \sep EEG \sep  Neural Networks \sep  EEGNet
\end{keyword}

\end{frontmatter}


\section{Introduction}
Artificial Neural Networks made one of the earliest significant impacts in the field of Brain-Computer Interfaces (BCI) when Schirrmeister et al. introduced Deep ConvNet and Shallow ConvNet in 2017 \cite{schirrmeister_deep_2017} for electroencephalographic (EEG) signal classification. Since then, neural networks have become one of the hottest topics in BCI literature. 

\begin{figure}[htp]
    \centering
    \includegraphics[width=\linewidth]{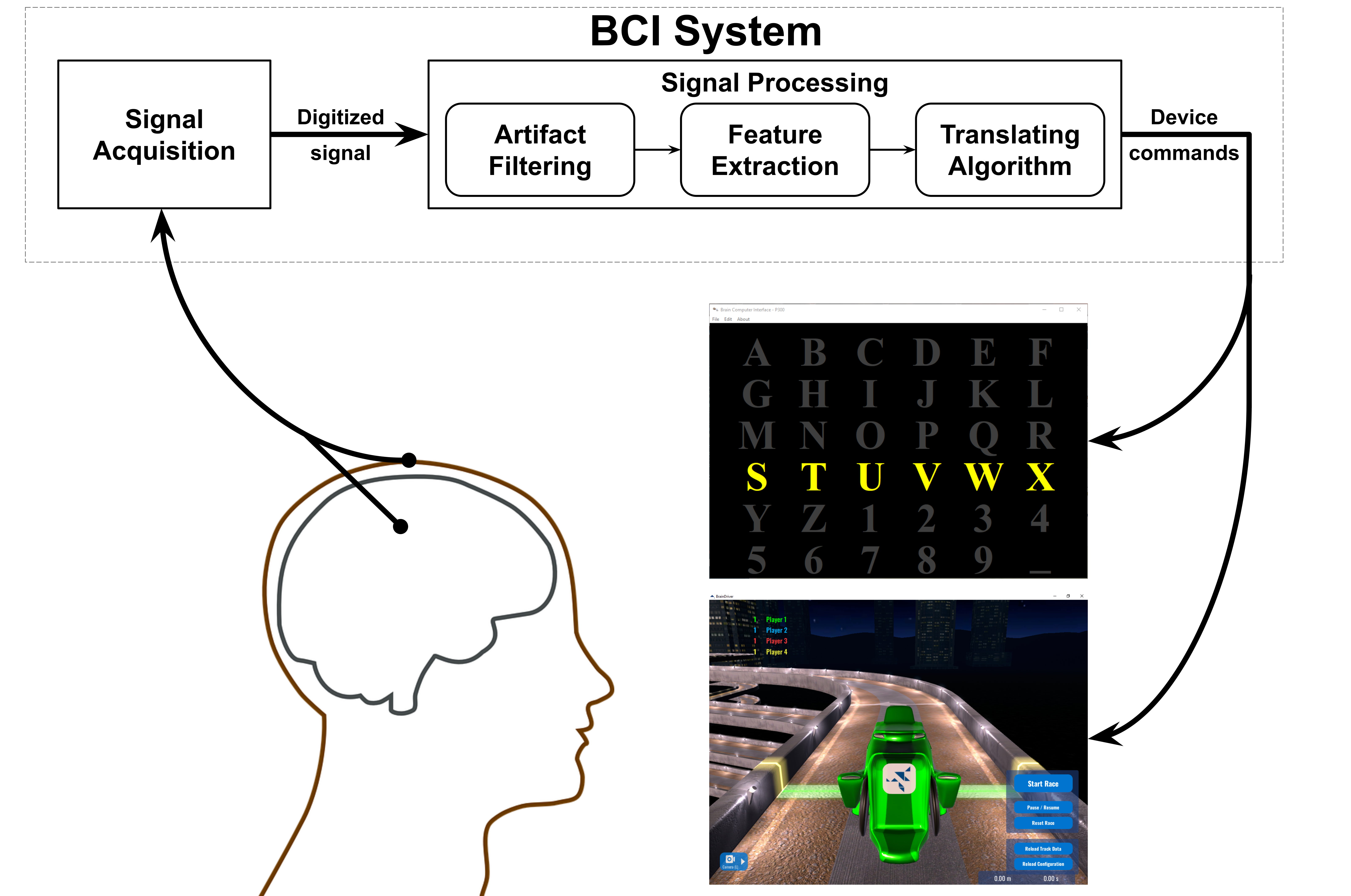}
    \caption{Components of a Brain-Computer Interface system}
    \label{fig:bci}
\end{figure}

BCIs are integrated systems that include software and hardware components. As it is presented by Wolpaw et al.\ \cite{wolpaw_braincomputer_2002}, these systems record bioelectrical signals from the brain, extract useful information from the EEG-noise mixture, and convert them to computer commands. EEG is defined as the postsynaptic membrane potential fluctuation of the neurons, recorded from the surface of the head. Figure \ref{fig:bci} presents the components of a BCI System.

If a new system is developed for motor imagery (MI) signal classification, it is often tested and compared with the previously published ones on one of the BCI Competition databases:  
\cite{blankertz_bci_2004, blankertz_bci_2006, sajda_data_2003, tangermann_review_2012}. 
However, these datasets contain records from less than or equal to 10 subjects. Other open-access databases have EEG records from more than 50 subjects but are avoided mainly by the researchers. One is the MI EEG dataset on PhysioNet \cite{goldberger_ary_l_physiobank_2000} recorded by the BCI2000 software \cite{schalk_bci2000_2004}, which includes EEG records from 109 subjects. The other one was recorded utilizing the OpenBMI toolbox \cite{lee_eeg_2019} and contained data from 52 subjects. Each subject in this dataset participated in two experimental days. In addition, we have recorded our dataset, which included 25 experiments from 9 subjects \cite{kollod_ttk_2022}. We hypothesize that databases with more than 20 experimental days are sufficient for BCI system comparison.

Next to the offline comparisons, the Cybathlon competition \cite{riener_cybathlon_2014} was introduced. It aimed to investigate the reliability of BCI systems working in real-time, out of the lab situation. 11 teams participated successfully in the BCI discipline of Cybathlon 2016 \cite{novak_benchmarking_2018}, and two published their concepts, training protocols, and BCI Systems after the competition. 
\cite{perdikis_cybathlon_2018, statthaler_cybathlon_2017}
As a continuation of this competition, the 2019 Cybathlon BCI Series and the 2020 Cybathlon Global Edition were organized, from which multiple teams shared their preparation and results. 
\cite{benaroch_long-term_2021, hehenberger_long-term_2021, korik_competing_2022, robinson_design_2021, tortora_neural_2022, turi_long_2021}

Before neural networks, scientists intended to investigate and develop hand-crafted feature extraction methods combined with simple classification algorithms. Blankertz et al.\ \cite{blankertz_non-invasive_2007} successfully used the Common Spatial Patterns (CSP) algorithm with Linear Discriminant Analysis (LDA) classifier to control the cursor in one dimension. Barachant et al.\ \cite{barachant_riemannian_2010} introduced Riemannian geometry for BCI with an LDA classifier to successfully classify EEG covariance matrices. Lotte and Guan  \cite{lotte_regularizing_2011} introduced a unifying theoretical framework for regularizing the CSP and compared it with 10 other regularized versions of the CSP algorithm. Another feature extraction algorithm, considering the CSP, is the Filter Bank Common Spatial Pattern (FBCSP) with a Naïve Bayesian Parzen Window classifier \cite{ang_filter_2012}, which was compared with the ConvNets \cite{lawhern_eegnet_2018, schirrmeister_deep_2017} on the BCI Competition IV 2a database. The winner of the BCI discipline of the Cybathlon competitions used the power spectral density of the EEG signals as a feature \cite{perdikis_cybathlon_2018, tortora_neural_2022} with a Gaussian classifier. 

With the introduction of the Deep and Shallow ConvNets, a new trend started in BCI development. The focus had been shifted from hand-crafted features to creating neural networks which not just classify the signal but also include the feature extraction step. Lawhern et al. \cite{lawhern_eegnet_2018} introduced the EEGNet, which was inspired by former neural networks that were designed for EEG signal processing, including MI-based BCIs \cite{sakhavi_parallel_2015, schirrmeister_deep_2017, sturm_interpretable_2016, tabar_novel_2017}. It was demonstrated that EEGNet makes a similar feature extraction as the FBCSP. This neural network inspired many scientists, which resulted in many improved versions of the EEGNet 
\cite{altaheri_physics-inform_2022, bria_sinc-based_2021, deng_advanced_2021, gao_convolutional_2022, huang_s-eegnet_2020, li_motor_2022, li_parallel_2022, liu_tcacnet_2022, ma_channel-mixing_2021, musallam_electroencephalography-based_2021, riyad_mi-eegnet_2021, riyad_novel_2021, roots_fusion_2020, yao_fb-eegnet_2022}, 
creating a whole family of the neural network. Other publications outside the EEGNet family \cite{dokur_classification_2021, fadel_multi-class_2020, fadel_chessboard_2020, han_classification_2022, jia_excellent_2023, jia_joint_2022, roy_adaptive_2022} 
highlight the importance of neural network-based BCI research.

Along with the development of neural networks, scientists started investigating the effect of transfer learning \cite{weiss_survey_2016}. This method aims to transfer knowledge between two domains and increase the classification’s accuracy. Khademi et al.\ \cite{khademi_transfer_2022} used a CNN-LSTM hybrid model, which was pretrained on the ILSVRC subset of the ImageNet dataset to classify MI EEG signals. This way, they aimed to transfer the knowledge of image classification and use it on spatial EEG images generated with continuous wavelet transformation method using complex Morlet mother wavelet. Another strategy is to utilize the entire EEG dataset and combine cross-subject and within-subject training, as presented in \cite{mattioli_1d_2022, roy_adaptive_2022, zhang_hybrid_2021}. 
In this case, the knowledge is granted from subjects not used in the test set of the neural network. The network is pretrained on data from all but one subject, as it would in a cross-subject training procedure. But then the data of the test subject is also split to train and test set, as in the within-subject, and the training part is used for fine-tuning the pretrained neural network. We selected the later version of transfer learning because it is architecture independent and aimed to use it after artifact filtering.

In this article, all the experiments were conducted on data purified from artifacts because eye and muscle movement activity can distort the EEG signal \cite{nolan_faster_2010}. Moreover, it was demonstrated that artifacts could successfully be used for BCI purposes \cite{noboa_development_2021}; however, in our perspective, an actual BCI does not depend on artifacts, only on pure brain waves. 

To reduce the computational time of the experiments, we have arbitrarily selected Shallow and Deep ConvNet \cite{schirrmeister_deep_2017} as predecessors of EEGNet, the EEGNet itself \cite{lawhern_eegnet_2018}, the EEGNet Fusion \cite{roots_fusion_2020}, and the MI-EEGNet \cite{riyad_mi-eegnet_2021} from the EEGNet family.

\section{Materials and Methods}
The databases and neural networks are presented in this section, with the experimental setups and concepts. The code used in this study is available under: \url{https://github.com/kolcs/bionic_apps}

\subsection{Databases}
In the following, we present the datasets which were used for the EEGNet family comparisons. The databases were processed in an “independent days” configuration, meaning that if a subject participated in an experiment multiple times on different experimental days, the data were handled as if it would have been recorded from various subjects. According to our knowledge, EEG data can greatly be varied by many factors, such as recording setup, period of the day, and mental state of the subjects. The latter could lead to poorer performance if the data is merged concerning the subjects. With the independent days configuration, we aimed to overcome this problem and extend the number of subjects to strengthen the results of the statistical analyses.

\subsubsection{Physionet}
The open-access database PhysioNet (Goldberger Ary L. et al., 2000) is a valuable resource for numerous physiological datasets. Among these datasets is the EEG Motor Movement/Imagery Dataset, which was captured by Schalk et al. \cite{schalk_bci2000_2004} utilizing the BCI2000 paradigm control program. For convenience, we will refer to this specific dataset as the Physionet database. It encompasses four MI EEG signals obtained from 109 individuals: Left Hand, Right Hand, Both Hands, and Both Legs. The MI periods are of 4-second duration and are interspersed with 4-second long Rest periods. The recordings were sampled at 160 Hz, over 64 channels, without using hardware filters. 

Four subjects out of the 109 were dropped from the database before the experiments. For subject 89 the labels were incorrect. In the case of subjects 88, 92, and 100 the timing was incorrect. The execution of the MI tasks and the resting phases were 5.125 and 1.375 seconds, respectively. Moreover, the sampling frequency was changed from 160 Hz to 128 Hz. Other articles, which use the Physionet database \cite{fan_bilinear_2020, mattioli_1d_2022, roots_fusion_2020}, also reported these problems

\subsubsection{Giga}
Lee et al.\ \cite{lee_eeg_2019} published an EEG dataset that included three paradigms: MI, event-related potential, and steady-state visually evoked potential. The experimental paradigms were conducted with the OpenMBI toolbox, custom written in MATLAB. We selected the files corresponding to the MI EEG paradigm from these three paradigms, which contains a 2-class classification problem, where the tasks are Left Hand and Right Hand movement imagination. The EEG signals were recorded with a 62-channeled BrainAmp amplifier system with a sampling rate of 1000 Hz. 54 subjects participated in the experiments, and each subject was present on two different experimental days. Therefore, concerning our independent days configuration, this dataset contains 108 subjects. To shrink the size of the raw EEG files, we resampled the data to a 500 Hz sampling frequency.

\subsubsection{BCI Competition IV 2a}
Tangermann et al.\ \cite{tangermann_review_2012} published the well-known and highly used BCI competition IV database, which includes 5 sub-datasets with different paradigms and challenges. This popular dataset is used as a standard in the BCI literature to compare the developed methods and algorithms. This article uses only the 2a sub-dataset, an MI dataset with Left Hand, Right Hand, Both Feet, and Tongue tasks. The EEG signals were recorded with a 250 Hz sampling frequency on 22 electrodes. The amplifier included a hardware bandpass filter between 0.5 and 100 Hz and a notch filter at 50 Hz to remove the powerline noise. 

This dataset was recorded with the help of 9 experimental subjects, and each subject participated in two different experimental days. Therefore, concerning the independent days configuration, this dataset contains 18 subjects.

\subsubsection{TTK}
The TTK database \cite{kollod_ttk_2022} was recorded in Research Centre for Natural Sciences (TTK, as a Hungarian abbreviation). A 64-channeled ActiChamp amplifier system (Brain Products GmbH, Gliching, Germany) was used to capture the EEG signals, which were operated with a 500 Hz sampling frequency.

The EEG signals were recorded with a custom designed MATLAB based, paradigm leader code called General Offline Paradigm (GoPar). GoPar was presented in the Supplementary Materials of \cite{kollod_closed_2023} and available at \url{https://github.com/kolcs/GoPar}. The paradigm of the Physionet database inspired this code. GoPar was designed to conduct multiple different MI paradigms with 4 tasks. In the case of the TTK dataset, these tasks were Left Hand, Right Hand, Left Foot, and Right Foot. The paradigm started with a one-minute-long eye-open session, followed by a one-minute-long eye-closed one as the initial task. These tasks aimed to get the subjects’ full attention, preparing them for the core part of the experiment, and served as a baseline. The paradigm continued with 2 warmup sessions where two out of the four MI tasks were selected and practiced overtly and covertly. The warmup sessions aimed to lead the subjects on how to execute MI tasks. After the warmup sessions, the pure MI sessions were followed with a randomized order of the four MI tasks. 

In total 25 experiments were conducted with 9 subjects. No hardware or software filters were applied.

\subsection{Signal processing}\label{sec:sig_proc}
As a first step, the EEG signals were filtered with a 5th-order Butterworth bandpass filter between 1 and 45 Hz. Then a customized FASTER algorithm \cite{nolan_faster_2010} was applied, as presented in \cite{kollod_closed_2023}, to remove artifacts corresponding to eye movement or muscle activity. The first step removed EEG channels which were assumed to be constantly noisy throughout an entire experiment, concerning the variance, the correlation, and the Hurst exponent. Secondly, epochs containing motions (chewing, yawning) were discarded by measuring the deviation from the channel average, the amplitude range, and variance parameters. In the third step eye related artifacts are filtered out utilizing independent component analysis. The final step filtered out EEG channels from epochs individually, which were still considered noisy, concerning the variance, median gradient, amplitude range, and channel deviation parameters. The fifth step of the original FASTER algorithm, which detected artifacts through subjects, was omitted, because we aimed our signal processing algorithm to be subject-specific.

The purified 4-second-long epochs were split into 2-second-long windows with 0.1-second shifts to enhance the number of samples. Then the signals were normalized using the standard scaling, where the mean of the data is set to zero and the standard deviation to one. These processed EEG windows were used to train and test the classifiers of the BCI System. 

In the case of within-subject classification, 5-fold cross-validation was conducted subject-wise, where the database was split on the epoch level to ensure that windows originating from the same epoch are exclusively used in either the train or test set. Approximately 10 \% of the data from the training set were used as a validation set, splitting it on the epoch level.

\subsection{Nerual Networks}
This section presents the used neural networks, our methods, and modifications concerning the original ones.

\subsubsection{Callbacks}
Under the training of neural networks, a modified early stopping and model-saving strategy were applied. The conventional early stopping approach \cite{prechelt_early_2012} focuses on monitoring the validation loss and halting the learning process when it increases to prevent the network from overfitting. Additionally, a patience parameter can be defined to determine the number of training epochs that should be waited for before the monitored value shows improvement again. We extended this strategy with an additional patience-like value called “give up”. This strategy aims to handle training situations where the validation loss increases above the initial training loss, and after a particular time, the neural network starts to learn; therefore, the validation loss decreases. The give up value defines how many training epochs should be waited until the validation loss reaches the initial amount of validation loss. The original patience value is activated if the initial loss has been reached under the give up limit. Otherwise, the training is finished. 

Our model saving strategy was designed to reflect on the modified early-stopping strategy. Until the initial validation loss was reached, model weights with the highest validation accuracy were saved. After reaching the initial validation loss, model weights were only saved if improvements were detected with respect to both the validation loss and validation accuracy. Before the test phase, the best model weights were restored. 

We conducted our experiments by setting the training epochs to 500, the give up value to 100, and the patience to 20.

\subsubsection{ConvNets}
For implementing the Deep and Shallow ConvNets the source code in \cite{lawhern_eegnet_2018} was used, which uses a few modified parameters concerning the originally published ones in \cite{schirrmeister_deep_2017}. No additional modifications were made concerning the architecture of the networks.

\subsubsection{EEGNets}
The networks of the EEGNet family, the EEGNet \cite{lawhern_eegnet_2018}, the EEGNet Fusion \cite{roots_fusion_2020}, and the MI-EEGNet \cite{riyad_mi-eegnet_2021} were all modified in such a way that they could be automatically used for databases with different sampling frequencies instead of setting up the input parameters manually. In the article of the EEGNet \cite{lawhern_eegnet_2018}, the authors explicitly said that the filter size of the first convolutional block should be half of the sampling frequency rate. Therefore, in our implementation, instead of directly defining the size of the kernel, it is calculated from the used signals sampling frequency. We followed this strategy in the case of the other two networks.

\subsection{Transfer learning}
Next to the subject-wise learning, we also investigated the effect of transfer learning. First, test subjects were selected as distinct groups of 10. The remaining subjects, called pre-train subjects, were used to set the initial optimal weights for the neural networks. A validation set was separated from the pre-train data to use it with our modified early stopping and model-saving strategy. When the pretraining phase converged, either reaching the maximum training epoch number or stopping it by the early stopping strategy, the best weights of the network were stored. For each test subject, 5-fold within-subject cross-validation was conducted as described in the third paragraph of \ref{sec:sig_proc}. Before each cross-validation step, the saved model’s weights were loaded, and the selected training set of the selected test subject was used as a fine-tuning data for the neural networks. Under the fine-tuning step, validation sets were again used with our early-stopping and model-saving strategies.

\subsection{EEGNet family comparison}
Many computational experiments were conducted on each database (Physionet, Giga, TTK, and BCI Competition IV 2a) to compare the neural networks from the EEGNet family (Shallow ConvNet, Deep ConvNet, EEGNet, EEGNet Fusion, MI-EEGNet). In the case of an experimental subject who participated in multiple experiments on different days, the data was handled as if multiple subjects would have participated instead, called independent days configuration. However, on the BCI Competition IV 2a dataset, we also conducted experiments where the data of a subject is combined, overseeing the date of the records because this way, the results are more comparable with previous BCI studies. We highlighted this experiment with the “merged subject data” words.

A within-subject and a transfer learning phase were conducted on each neural network database. The results of the cross-validations were collected, and the normality test was used to select a proper statistical test: the t-test or Wilcoxon for normally or not normally distributed accuracy levels, respectively. The received p-values were adjusted with Bonferroni correction. The significance level was preset to 0.05.

We aimed to rank the neural networks; therefore, two additional metrics were introduced next to the pure accuracy comparison. These metrics were investigated on the independent days configured databases. The first metric is the accuracy improvement of the EEGNet family with respect to the chance level. One advantage of this metric is that it can be used on databases with different class numbers. This metric was calculated and averaged for both within-subject and transfer learning. 

The second metric focused on the effect of transfer learning, which was investigated by comparing the results of within-subject classification with the transfer learning classification. The difference between the two methods was calculated for each database concerning independent days.

\subsection{Significance investigation of databases}
The number and quality of significant differences were investigated to quantitatively measure our hypothesis that databases with higher than 20 experimental days are sufficient only for BCI system comparison. For each database configuration, we calculated two numbers. The sum of significance levels, where the significance levels were categorized as it, is presented in Table \ref{tb:sig_level}.\ and the count of significant differences. We correlated these numbers with the number of subjects in the databases.

\begin{table}[htp]
\centering
\caption{Level of a significance tests}
\begin{tabular}{cr}
\textbf{Level} & \textbf{p-value range}             \\
\hline
1              & 1e-2 \textless{ }p \textless{}= 5e-2 \\
2              & 1e-3 \textless{ }p \textless{}= 1e-2 \\
3              & 1e-4 \textless{ }p \textless{}= 1e-3 \\
4              & p \textless{}= 1e-4               
\end{tabular}
\label{tb:sig_level}
\end{table}

\section{Results}
\begin{figure*}[htp]
    \thispagestyle{empty}
    \centering
    \vspace{-1.1in}
    \centerline{
        \includegraphics[width=1.3\textwidth]{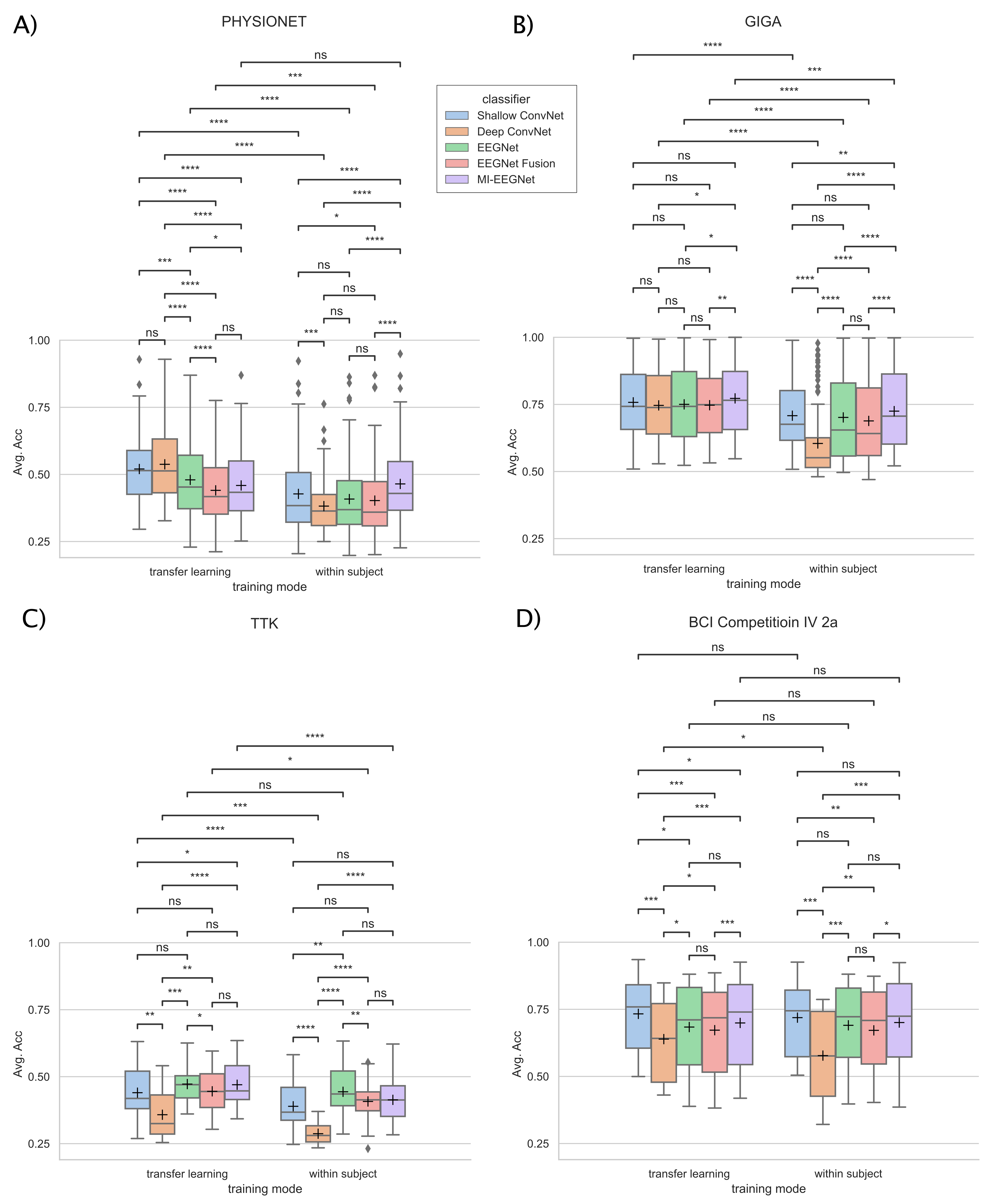}}
    \caption{EEGNet family comparison on 4 databases handling the datasets in independent days configuration. \\The p-value annotation legend is the following: ns:5e-2 < p; *: 1e-2 < p <= 5e-2; **: 1e-3 < p <= 1e-2; ***: 1e-4 < p <= 1e-3; ****: p <= 1e-4.  The mean of the data is presented with the ‘+’ symbol.}
    \label{fig:grouped}
\end{figure*}

After receiving the 5-fold cross-validated accuracy levels for all the combinations of the 4 databases, 5 neural networks, and 2 learning methods (within-subject and transfer learning), the normality tests showed a not normal distribution. Therefore, the Wilcoxon statistical test was used with Bonferroni correction for significance analysis. The results are presented in Figure \ref{fig:grouped}. In general, it can be concluded that transfer learning significantly improved the results at all the databases except the BCI Competition 2a.
\newpage

\thispagestyle{empty}
\clearpage

On the Physionet (Figure \ref{fig:grouped}A), in the case of within-subject classification, the MI-EEGNet significantly reached the highest accuracy (0.4646) compared to the other methods, while in the case of transfer learning, the Deep ConvNet showed the significantly highest performance (0.5377).

On the Giga database (Figure \ref{fig:grouped}B), the MI-EEGNet reached the highest accuracies, 0.725 and 0.7724, in the case of within-subject and transfer learning, respectively. This network significantly outperformed the others, except compared to the Shallow ConvNet in transfer learning mode.

Analyzing the results from the TTK dataset (Figure \ref{fig:grouped}C), it can be concluded that the EEGNet reached the highest accuracies: 0.4437 and 0.4724 in the case of within-subject and transfer learning, respectively. These results were significantly higher than the other networks except for the MI-EEGNet.

\begin{figure}[htp]
    \centering
    \includegraphics[width=.75\linewidth]{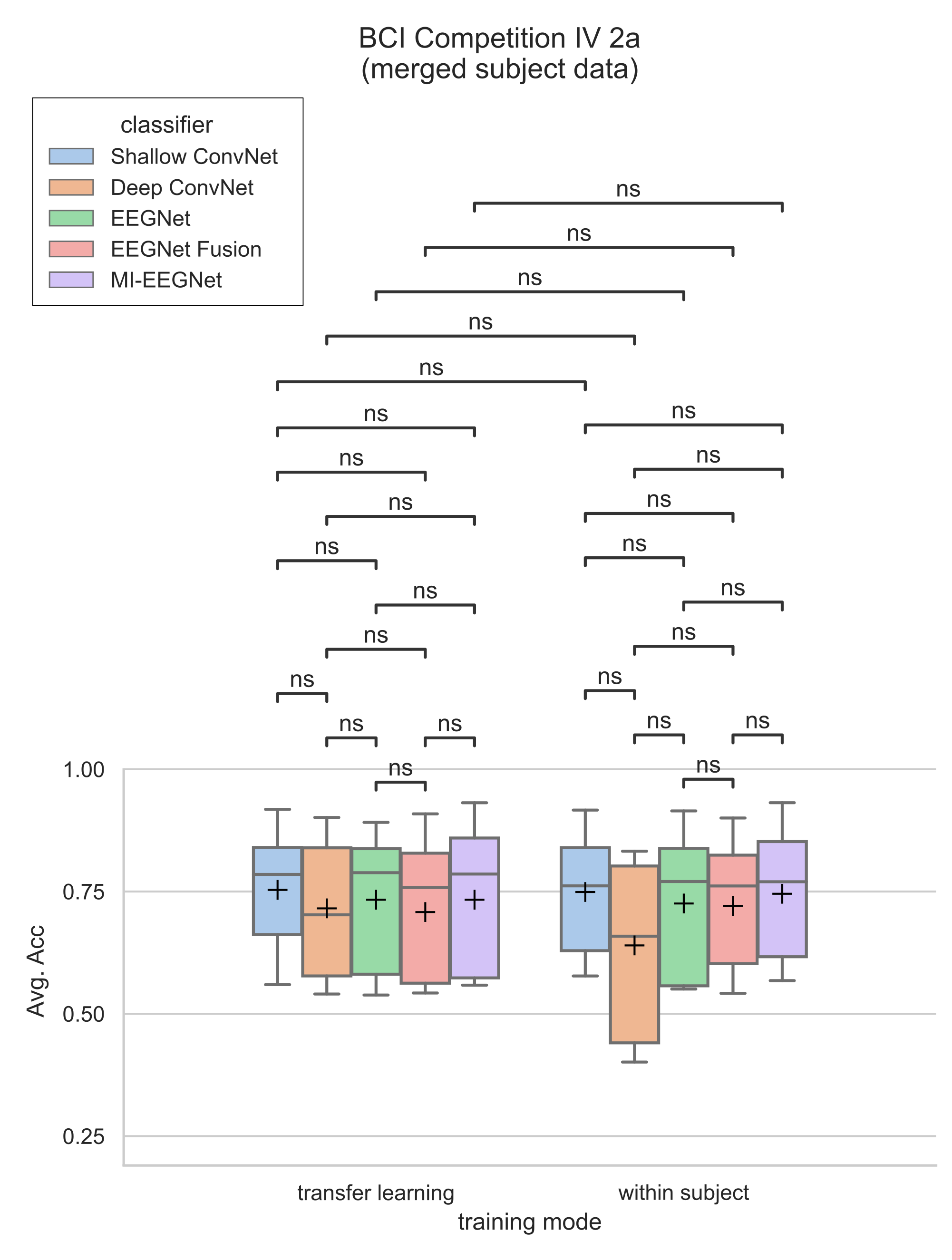}
    \caption{EEGNet family comparison on BCI Competition IV 2a. The p-value annotation legend is the following: ns:5e-2 < p; *: 1e-2 < p <= 5e-2; **: 1e-3 < p <= 1e-2; ***: 1e-4 < p <= 1e-3; ****: p <= 1e-4.  The mean of the data is presented with the ‘+’ symbol.}
    \label{fig:merged}
\end{figure}

In the case of the BCI Competition IV 2a dataset, handled as independent days (Figure \ref{fig:grouped}D), the Shallow ConvNet reached 0.719 and 0.733 accuracies in the case of within-subject and transfer learning, respectively. In the case of transfer learning, it was significant compared to the other neural networks. However, concerning the within-subject classification, the results were comparable with the EEGNet and the MI-EEGNet. On the other hand, when the data corresponding to one subject was merged, ignoring the experimental days, the Shallow ConvNet reached again the highest accuracies, 0.749 and 0.7533 in the case of within-subject and transfer learning, respectively; however, the difference between the neural network was insignificant. 

To establish a hierarchy among the neural networks, we analyzed the accuracy improvement of the EEGNet family relative to the chance level. Table \ref{tb:avg_dist} displays the ranking of these networks based on their training modes. Among all the databases in an independent days setup, MI-EEGNet displayed the most significant average improvement in within-subject classification. On the other hand, the Shallow ConvNet outperformed the other networks for transfer learning.

\begin{table*}[htp]
\centering
\caption{Classification Improvements by transfer learning on databases with independetn day configuration}
\centerline{\begin{tabular}{clrrrrr}
\textbf{Rank} & \textbf{Neural Networks} & \textbf{Physionet} & \textbf{Giga} & \textbf{TTK} & 
    \begin{tabular}[c]{@{}r@{}}\textbf{BCI Comp}\\ \textbf{IV 2a}\end{tabular} & \textbf{Avg. impr.} \\ \hline
1 & Deep ConvNet & 0.1557 & 0.1418 & 0.0708 & 0.0614 & 0.1075 \\
2 & Shallow ConvNet & 0.0928 & 0.0497 & 0.0509 & 0.0141 & 0.0519 \\
3 & EEGNet & 0.0716 & 0.0487 & 0.0288 & -0.0065 & 0.0357 \\
4 & EEGNet Fusion & 0.0381 & 0.0586 & 0.0379 & 0.0007 & 0.0338 \\
5 & MI-EEGNet & -0.0058 & 0.0475 & 0.0564 & -0.0015 & 0.0241
\end{tabular}}
\label{tb:transfer_learning}
\end{table*}

\begin{table}[htp]
\centering
\caption{Ranking the performance of neural networks on databases concerning independent days.}
\begin{tabular}{lclc}
\textbf{} & \textbf{Rank} & \textbf{Classifier} & \textbf{\begin{tabular}[c]{@{}c@{}}Avg. Acc. impr. \\ from chance level\end{tabular}} \\ \hline
\multirow{5}{*}{\begin{tabular}[c]{@{}l@{}}Within\\ subject\end{tabular}} & 1 & MI-EEGNet & 0.2306 \\
 & 2 & Shallow ConvNet & 0.2071 \\
 & 3 & EEGNet & 0.1997 \\
 & 4 & EEGNet Fusion & 0.1871 \\
 & 5 & Deep ConvNet & 0.1249 \\ \hline
\multirow{5}{*}{\begin{tabular}[c]{@{}l@{}}Transfer\\ learning\end{tabular}} & 1 & Shallow ConvNet & 0.2721 \\
 & 2 & Deep ConvNet & 0.2598 \\
 & 3 & MI-EEGNet & 0.2537 \\
 & 4 & EEGNet & 0.2521 \\
 & 5 & EEGNet Fusion & 0.2312
\end{tabular}
\label{tb:avg_dist}
\end{table}

Another factor we considered was the enhancement of neural networks through transfer learning, which we have presented in Table \ref{tb:transfer_learning}. Deep ConvNet showed the most substantial improvement, achieving results 0.1 higher than within-subject classification mode on average. Whereas Shallow ConvNet, ranked first in transfer learning performance, could only improve by 0.05 compared to within-subject classification.

Finally, the databases were ranked concerning the number of significant differences. Table \ref{tb:signif} presents the sum of significance ranges (corresponding to the number of stars on the figures) and the count of significant differences with the number of subjects in the databases. The sum of the significance ranges and the number of subjects in the databases were found to be strongly correlated, r(3) = 0.7709, but insignificant (p-value 0.127014 > 0.05).

\begin{table}[htp]
\centering
\caption{Significance investigation}
\begin{tabular}{cccc}
\textbf{} & \multicolumn{2}{c}{\textbf{Significance level}} & \textbf{} \\
\textbf{Database} & \textbf{Sum} & \textbf{Count} & \textbf{Subjects} \\ \hline
Physionet & 63 & 18 & 105 \\
Giga & 49 & 15 & 108 \\
TTK & 45 & 16 & 25 \\
BCI Comp IV 2a & 31 & 15 & 18 \\
\begin{tabular}[c]{@{}c@{}}BCI Comp IV 2a –\\ merged subject data\end{tabular} & 0 & 0 & 9
\end{tabular}
\label{tb:signif}
\end{table}

\section{Discussion}
Most of the articles presenting MI EEG signal classification with artificial neural networks from the EEGNet family are presenting and comparing their results on one of the BCI competition databases. This article aimed to highlight that datasets with many subjects are required for statistically significant comparisons. Therefore, we compared 5 neural networks from the EEGNet family on 4 databases which include various subjects. Concerning the datasets, we introduced the independent day configuration, where the data of a subject who participated in multiple experimental days counted as if multiple subjects participated instead. With this configuration, we aimed to extend the number of experiments and increase the comparisons’ significance level. All four databases, namely the BCI competition IV 2a database \cite{tangermann_review_2012}, the Physionet \cite{goldberger_ary_l_physiobank_2000, schalk_bci2000_2004}, the Giga \cite{lee_eeg_2019}, and our TTK dataset \cite{kollod_ttk_2022}, were used in this configuration. In the case of Physionet, the authors claim that the experiments were conducted with 109 volunteers; therefore, the independent subject configuration is irrelevant here. Concerning the BCI competition IV 2a database, we conducted another experiment where the data of a subject was combined into one (“merged subject data”). Width this configuration, we aimed to conduct classifications comparable to other reports. In addition, we used these results to test our hypothesis about the correlation between the number of subjects in the database and the number of significant comparisons (Table \ref{tb:signif}). The correlation between the number of subjects and our significance metric was strong but insignificant. However, Table \ref{tb:signif} presents that a database with 9 subjects is insufficient for significance testing. Therefore, we suggest that databases with many subjects, such as the Physionet or the Giga datasets, should be used to compare BCI systems. To further investigate our hypothesis, additional open-access MI EEG databases are required.

We also would like to highlight that in our experiments artifact filtered EEG data were used, unlike the articles about the investigated neural networks 
\cite{lawhern_eegnet_2018, riyad_mi-eegnet_2021, roots_fusion_2020, schirrmeister_deep_2017}. They included bandpass filtering and standardization before the classification. In our signal processing step, next to the 5th ordered bandpass Butterworth filter from 1 to 45 Hz, we utilized the FASTER \cite{nolan_faster_2010} artifact removal algorithm to detect and remove artifacts corresponding to eye movement and muscle activity. This is highly important; otherwise, it may happen that instead of pure EEG signals, artifacts are classified. It was demonstrated in \cite{noboa_development_2021} that electromyography could be successfully used for BCI purposes.

Most articles that investigate the effect of transfer learning are presented using datasets without artifact filtering 
\cite{kant_cwt_2020, khademi_transfer_2022, mattioli_1d_2022, roy_adaptive_2022, zhang_hybrid_2021}. We showed that on databases with many subjects (Physionet and Giga), even after artifact filtering, the transfer learning significantly improves the accuracy of the neural networks compared to the within-subject classification (Figure \ref{fig:grouped} A \& B). We also showed that Deep ConvNet could gain the most from the transfer learning considering all the databases (Table \ref{tb:transfer_learning}). On the other hand, the Shallow ConvNet reached the highest results concerning our “improvement from the chance level” metric on all transfer learning trained neural networks (Table \ref{tb:avg_dist}). Nevertheless, the difference between the ConvNets is insignificant concerning the Physionet and the Giga database (Figure \ref{fig:grouped} A \& B). 

Our results highlight that various aspects should be considered for a good ranking between neural networks. Using unfiltered datasets with few numbers of subjects and considering only the accuracy differences between networks may lead to a vague conclusion.

In future work, it would be interesting to conduct transfer learning by utilizing data from multiple databases. However, one must overcome the problem that different datasets are recorded with different EEG amplifiers; therefore, the position, number of electrodes, and sampling frequency could differ.

\section{Conclusion}
In this paper, we critically compared neural networks from the EEGNet family, namely the Shallow ConvNet, the Deep ConvNet, the EEGNet, the EEGNet Fusion. and the MI-EEGNet, on MI EEG signal classification tasks. The comparisons were conducted utilizing the BCI competition IV 2a database and the Giga and the Physionet, which include many subjects. In addition, we also used our TTK dataset. Within-subject and transfer learning classification were conducted on each database configuration and neural network combination. All the received results were 5-fold cross-validated. The classification results were acquired after cleaning the raw signals from artifacts with the FASTER algorithm. 

To the best of our knowledge, there were no articles published before that compare neural networks from the EEGNet family on artifact filtered databases, which include high numbers of subjects (>20), and in addition, the results are also cross-validated. We also demonstrated that transfer learning could improve the classification results, even on artifact-filtered MI EEG data. To rank the neural networks, we introduced two metrics. The first considered the neural networks’ accuracy improvement from the chance level, while the second investigated the classification improvements by transfer learning. These metrics showed that Shallow ConvNet and Deep ConvNet could perform better than the later published networks from the EEGNet family. Finally, we showed that databases with few numbers of subjects ($\leq$10) are not sufficient for comparing BCI systems statistically.

\section*{CRediT Authorship contribution statement}
\noindent
\textbf{Csaba Köllőd}: Conceptualization, Methodology, Experiments, Software, Statistics, Visualization, Writing – original draft\\
\textbf{András Adolf}: Software - FASTER algorithm implementation, Experiments, Writing - review \& editing\\
\textbf{Gergely Márton}: Experiments, Writing - review \& editing\\
\textbf{István Ulbert}: Supervision, Project administration, Writing - review \& editing

\section*{Declaration of Competing Interest}
The authors declare that they have no known competing financial interests or personal relationships that could have appeared to influence the work reported in this paper.

\section*{Acknowledgement}
We thank our Pilots for their weekly availability and participation in the experiments. We are also grateful for their regular feedback about the system.

Prepared with the professional support of the Doctoral Student Scholarship Program of the Co-operative Doctoral Program of the Ministry of Innovation and Technology financed from the National Research, Development and Innovation Fund.

\section*{Data availability statement}
Databases and source codes are available under the following links.
\newline

\textbf{Source codes:}
\begin{itemize}
    \item Signal processing and classification framework – \url{https://github.com/kolcs/bionic_apps}
	\item Paradigm Handler – \url{https://github.com/kolcs/GoPar}
\end{itemize}

\textbf{Datasets:}
\begin{itemize}
    \item Physionet – \url{https://physionet.org/content/eegmmidb/1.0.0/}
    \item Giga – \url{http://gigadb.org/dataset/100542}
    \item BCI Competition IV – \url{https://www.bbci.de/competition/iv/}
    \item TTK – \url{https://hdl.handle.net/21.15109/CONCORDA/UOQQVK}
\end{itemize}

\section*{Supplementary Materials}
Excel tables about database accuracies:
\href{https://ppke-my.sharepoint.com/:x:/g/personal/uwr78z_ad_ppke_hu/EaVuvCSpv_9MqKcvYIzkIisBEUvOoFAQK4sgEx7_pNgK-g?e=rHJbsQ}{Supplementary Materials.xlsx}



\printbibliography
\end{document}